\documentclass[ALICE,manyauthors]{cernphprep}

\usepackage[bookmarks]{hyperref}

\usepackage{lineno}

\def\snn{\mbox{$\sqrt{s_{_{\rm NN}}}$}}

\def\pt{p_{\rm{T}}}

\def\psin{\Psi_{\rm n}}

\usepackage{color}
\definecolor{dgreen}{cmyk}{1.,0.,1.,0.2}        
\definecolor{orange}{cmyk}{0.,0.353,1.,0.}    

\usepackage{epstopdf}
\begin{document}%
%
%
\begin{titlepage}
\PHnumber{2012-142}                 
\PHdate{24 May 2012}              
%

\title{Anisotropic flow of charged hadrons, pions and (anti-)protons
measured at high transverse momentum in Pb-Pb collisions at $\snn=2.76$~TeV}
\ShortTitle{Anisotropic flow at high transverse momentum}   

\Collaboration{ALICE Collaboration%
         \thanks{See Appendix~\ref{app:collab} for the list of collaboration
                      members}}
\ShortAuthor{ALICE Collaboration}      


\begin{abstract}
The elliptic, $v_2$, triangular, $v_3$, and quadrangular, $v_4$,
azimuthal anisotropic flow coefficients are measured for unidentified
charged particles, pions and (anti-)protons in Pb-Pb collisions at
$\snn = 2.76$~TeV with the ALICE detector at the Large Hadron
Collider. Results obtained with the event plane and four-particle
cumulant methods are reported for the pseudo-rapidity range
$|\eta|<0.8$ at different collision centralities and as a function of
transverse momentum, $\pt$, out to $\pt=20$~GeV/$c$. The observed
non-zero elliptic and triangular flow depends only weakly on
transverse momentum for $\pt>8$~GeV/$c$. The small $\pt$ dependence of
the difference between elliptic flow results obtained from the event plane and
four-particle cumulant methods suggests a common origin of flow
fluctuations up to $\pt=8$~GeV/$c$. The magnitude of the (anti-)proton
elliptic and triangular flow is larger than that of pions out to at
least $\pt=8$~GeV/$c$ indicating that the particle type dependence
persists out to high $\pt$.
\end{abstract}
\end{titlepage}
\setcounter{page}{2}

The goal of ultra-relativistic nucleus-nucleus collisions is to study
nuclear matter under extreme conditions. For non-central collisions, 
in the plane perpendicular to the beam direction, the geometrical
overlap region, where the highly Lorentz contracted nuclei
intersect and where the initial interactions occur, is 
azimuthally anisotropic. This initial spatial asymmetry is
converted via interactions into an anisotropy in momentum space, a
phenomenon referred to as transverse anisotropic flow (for a review
see~\cite{Voloshin:2008dg}). Anisotropic flow has become a key
observable for the characterization of the properties and the
evolution of the system created in a nucleus-nucleus collision.

Identified particle anisotropic flow provides valuable information on 
the particle production mechanism in different transverse momentum, $\pt$, 
regions~\cite{Voloshin:2008dg}. For $\pt < 2-3$~GeV/$c$, the flow 
pattern of different particle species is qualitatively described by 
hydrodynamic model calculations~\cite{Snellings:2006qw}. At intermediate 
$\pt$, $3 < \pt < 6$~GeV/$c$, the observed flow of the baryons 
is larger than that of the mesons~\cite{Abelev:2007rw,Adare:2012vq}. For 
$\pt \gtrsim 8$ GeV/$c$, the fragmentation of high-energy partons, resulting 
from initial hard scatterings, is expected to play the dominant role. While 
traversing the hot and dense 
matter these partons experience collisional and radiative energy 
loss~\cite{Gyulassy:1990ye,Wang:1991xy}, which are strongly dependent on the 
thickness of the created medium~\cite{Gyulassy:1993hr}. In the azimuthally 
asymmetric system, the energy loss depends on the azimuthal emission angle 
of the parton, which leads to an azimuthal anisotropy in particle production 
at high $\pt$~\cite{Snellings:1999gq,Wang:2000fq}.

The magnitude of the anisotropic flow is characterized by the coefficients in 
the Fourier expansion of the azimuthal distribution of particles with respect 
to the collision symmetry plane~\cite{Voloshin:1994mz,Poskanzer:1998yz}:
\begin{equation}
 v_n(\pt,\eta) = \langle \cos[n(\phi-\psin)] \rangle,
\end{equation}
where $\pt$, $\eta$, and $\phi$ are the particle's transverse
momentum, pseudo-rapidity, and the azimuthal angle, respectively, and
$\psin$ is the $n$-th harmonic symmetry plane angle. For a smooth
matter distribution in the colliding nuclei, the symmetry planes of
all harmonics coincide with the reaction plane defined by the beam
direction and the impact parameter, the vector connecting the centers
of the two colliding nuclei at closest approach. In this case, for
particles produced at midrapidity, all odd Fourier coefficients are
zero by symmetry. Due to event-by-event fluctuations of the positions
of the participating nucleons inside the nuclei, the shape of the
initial energy density of the heavy-ion collision in general is not
symmetric with respect to the reaction plane, and the $\psin$ may
deviate from the reaction plane.  This gives rise to non-zero odd
harmonic
coefficients~\cite{Mishra:2007tw,Mishra:2008dm,Takahashi:2009na,Alver:2010gr,Alver:2010dn,Teaney:2010vd,Luzum:2010sp},
and contributes to the difference in flow coefficients calculated from
two- or multi-particle azimuthal correlations, and also to the
difference in $v_n$ measured with respect to different harmonic
symmetry planes.

Large elliptic flow, $v_2$, and significant triangular flow, $v_3$, were 
observed at the Relativistic Heavy Ion Collider 
(RHIC)~\cite{Ackermann:2000tr,Adler:2003kt,Adare:2011tg} and at the Large 
Hadron Collider (LHC)~\cite{Aamodt:2010pa, Aamodt:2011hh, ATLAS:2011ah, Chatrchyan:2012wg, Aad:2012bu, Chatrchyan:2012ta, Chatrchyan:2012}. In this paper we 
present the measurement of unidentified charged particle anisotropic flow out 
to $\pt=20$~GeV/$c$, and for protons and charged pions~\footnote{In this 
analysis we do not differentiate between particle and antiparticle.} out to 
$\pt=16$~GeV/$c$. We also present unidentified charged particle 
quadrangular flow, $v_4$, measured with respect to the second ($\Psi_{2}$) 
and fourth ($\Psi_{4}$) harmonic symmetry planes. 

The data sample recorded by ALICE during the 2010 heavy-ion run at the
LHC is used for this analysis. Detailed descriptions of the ALICE
detector can be found
in~\cite{Aamodt:2008zz,Carminati:2004fp,Alessandro:2006yt}. The Time
Projection Chamber (TPC) was used to reconstruct charged particle
tracks and measure their momenta with full azimuthal coverage in the
pseudo-rapidity range $|\eta|<0.8$, and for particle identification
via the specific ionization energy loss, $\mathrm{d}E/\mathrm{d}x$, in
the transverse momentum region $\pt > 3$ GeV/$c$. Two scintillator
arrays (VZERO) which cover the pseudo-rapidity ranges $-3.7<\eta<-1.7$
and $2.8<\eta<5.1$ were used for triggering, and the determination of
centrality~\cite{Aamodt:2010cz} and symmetry planes. The trigger
conditions and the event selection criteria are identical to those
described in~\cite{Aamodt:2010pa, Aamodt:2011hh, Aamodt:2010cz}. 
Approximately $10^7$ minimum-bias Pb-Pb events with
a reconstructed primary vertex within $\pm 10$ cm from the nominal
interaction point in the beam direction are used for this
analysis. Charged particles reconstructed in the TPC in $|\eta|<0.8$
and $0.2<\pt<20$ GeV/$c$ were selected. The charged track quality cuts
described in~\cite{Aamodt:2010pa} were applied to minimize
contamination from secondary charged particles and fake tracks. The
charged particle track reconstruction efficiency and contamination
were estimated from {\sc HIJING} Monte Carlo
simulations~\cite{refHIJING} combined with a {\sc 
GEANT3}~\cite{geant3} detector model, and found to be independent of
the collision centrality. The reconstruction efficiency increases from
70\% to 80\% for particles with $0.2<\pt<1$~GeV/$c$ and remains
constant at $80 \pm 5$\% for $\pt>1$~GeV/$c$. The estimated
contamination by secondary charged particles from weak decays and
photon conversions is less than 6\% at $\pt=0.2$~GeV/$c$ and falls
below 1\% for $\pt>1$~GeV/$c$.

The selection of pions and protons at $\pt>3$ GeV/$c$ is based on the
measurement of the $\mathrm{d}E/\mathrm{d}x$ in the TPC, following the
procedure described in~\cite{Aamodt:2011zj}. Enriched pion (proton)
samples are obtained by selecting tracks from the upper (lower) part
of the expected pion (proton) $\mathrm{d}E/\mathrm{d}x$
distribution. For example, protons were typically selected, depending
on their momentum, in the range from 0 to $-3\sigma$ or from
$-1.5\sigma$ to $-4.5\sigma$ around their nominal value in
$\mathrm{d}E/\mathrm{d}x$, where $\sigma$ is the energy loss
resolution. Note that $\mathrm{d}E/\mathrm{d}x$ of pions is larger
than that of protons in the $\pt$ range used for this study. The track
selection criteria have been adjusted to keep the contamination by
other particle species below 1\% for pions and below 15\% for
protons. The pion and proton $v_2$ and $v_3$ are not corrected for
this contamination. The systematic uncertainties in $v_2$ and $v_3$
related to the purity of the pion and proton samples are 2\% for $\pt
< 8$ GeV/$c$ and 10\% for $\pt \ge 8$ GeV/$c$.

The flow coefficients $v_n$ are measured using the event plane method 
($v_n\{\rm{EP}\}$~\cite{Voloshin:2008dg}) and the four-particle cumulant 
technique ($v_n\{4\}$~\cite{Bilandzic:2010jr}), which have different 
sensitivity to flow fluctuations and correlations unrelated to the 
azimuthal asymmetry in the initial geometry (``non-flow''). The non-flow 
contribution to $v_n\{4\}$ is estimated to be negligible from analytic 
calculations and Monte Carlo 
simulations~\cite{Adler:2002pu, Pruthi:2011eq, Bilandzic:2012th}. The 
contribution from flow fluctuations was shown to be negative for 
$v_n\{4\}$ and positive for $v_n\{\rm{EP}\}$~\cite{Voloshin:2008dg}. 

\begin{figure}[t]  
  \begin{center}
    \includegraphics[keepaspectratio, width=0.97\columnwidth]{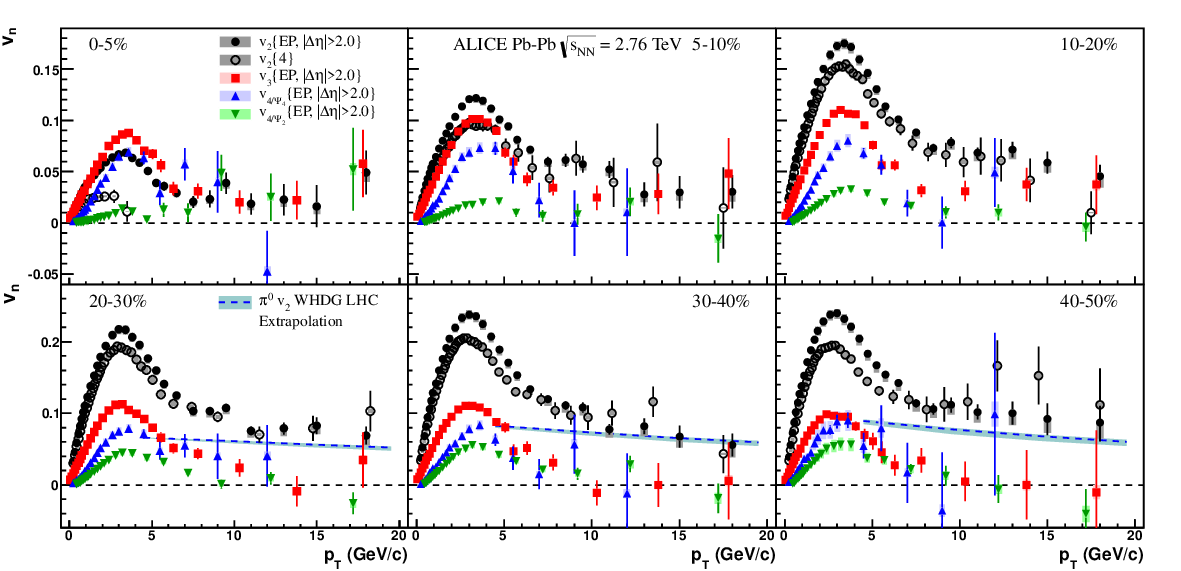}
    \caption{(color online) $v_2$, $v_3$, and $v_4$ measured for
      unidentified charged particles as a function of transverse
      momentum for various centrality classes. The dashed line
      represents the WHDG model calculations for neutral pions
      $v_2$~\cite{Horowitz:2011cv} extrapolated to the LHC collision
      energy. For clarity, the markers for $v_3$ and
      $v_{4/\Psi_{2}}$ results are slightly shifted along
      the horizontal axis. Error bars (shaded boxes) represent the 
      statistical (systematic) uncertainties.}
    \label{fig:vn}
  \end{center}
\end{figure}

The orientation of the symmetry 
planes $\psin$ is reconstructed from the azimuthal distribution of hits 
measured by the VZERO scintillators. The large gap in pseudo-rapidity between 
the charged particles in the TPC 
and those in the VZERO detectors greatly suppresses non-flow 
contributions to the measured $v_n\{\rm{EP}\}$. Assuming that there is no 
anisotropic flow in pp collisions, the non-flow contributions can 
be estimated by comparing the azimuthal correlations measured in heavy-ion 
collisions to those in pp. It was observed that the two-particle 
azimuthal correlations in pp and the most peripheral Au-Au 
collisions at $\sqrt{s_{\rm{NN}}}=0.2$~TeV are very similar~\cite{Adams:2004wz}, 
which suggests that non-flow dominates correlations in the centrality range 
80-90\%. The systematic uncertainty from the remaining non-flow, 
$\delta_{n}^{\rm cent}$, in the measured $v_n\{\rm{EP}\}$ coefficients was 
estimated based on the equation:
\begin{equation}
\delta_{n}^{\rm cent} = v_{n}^{80-90\%}\sqrt{\frac{M^{80-90\%}}{M^{\rm cent}}},
\label{eq:nonflow}
\end{equation}
where $v_{n}^{80-90\%}$ and $M^{80-90\%}$ are the magnitude
of $v_n$ and average multiplicity for the centrality range 80-90\%,
respectively, and $M^{\rm cent}$ is the average multiplicity in a
given centrality class. The non-flow increases with $\pt$ and from
central to peripheral collisions. For example, the non-flow
contributions to $v_2$ in 5-10\% (40-50\%) most central collisions are
about 1\% (2\%) at $\pt= 1$ GeV/$c$ and reach up to 10\% (12\%) for
$\pt> 10$ GeV/$c$. Other sources of systematic uncertainties were
evaluated from the variation of the results with different cuts on the
reconstructed collision vertex and the centrality estimated from the
charged particle multiplicity measured in the TPC and VZERO
detectors. Changes due to variations of the track selection criteria
and the difference of the results obtained using only positively or
negatively charged particles were considered as a part of the
systematic error. The difference in the extracted coefficients using
one or the other of the two VZERO detectors was found to be below 1\%
for $v_2$ and $v_3$, and below 5\% for $v_4$ over the measured region
of transverse momentum. The combined results from correlations with
both VZERO detectors are denoted as $v_n\{\rm{EP},|\Delta\eta|>2.0\}$
in the following. The contributions from all sources were added in
quadrature as an estimate of the total systematic uncertainty. The
resulting systematic uncertainties in $v_2$ are 3\% for $0.9<\pt<1$
GeV/$c$ and $^{+3}_{-11}$\% ($^{+3}_{-12}$\%) for $9<\pt<10$ GeV/$c$
in the 5-10\% (40-50\%) centrality class. The resulting systematic
uncertainties in $v_3$ are 3\% for $0.9<\pt<1$ GeV/$c$ and increase to
6\% (10\%) for $7<\pt<9$ GeV/$c$ for centrality 5-10\% (40-50\%). We
assign an 8\% (16\%) systematic uncertainty to $v_4$ for $0.9<\pt<1$
GeV/$c$ in the 5-10\% (40-50\%) centrality class, while for $\pt>6$
GeV/$c$ the systematic uncertainty is dominated by non-flow contributions.

\begin{figure}[t]   
  \begin{center}
  \includegraphics[keepaspectratio, width=0.7\columnwidth]{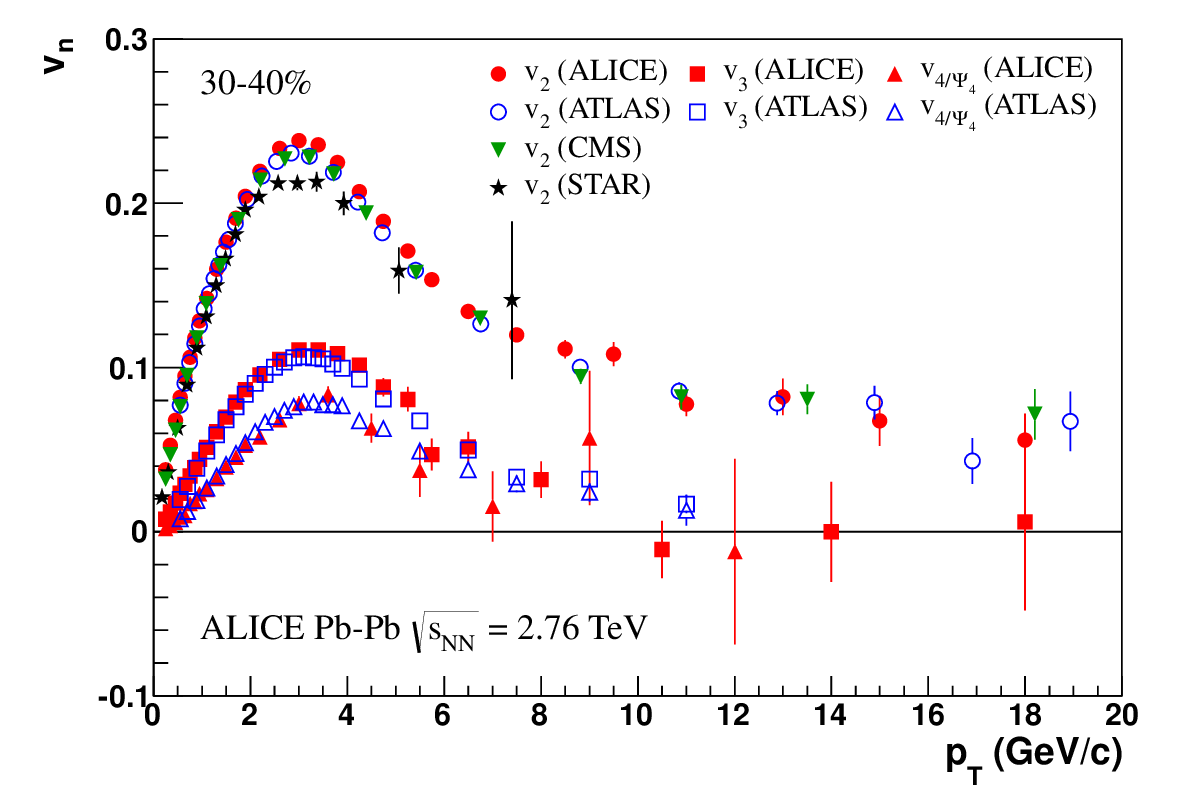}
  \caption{(color online) Comparison of the ALICE results on
      $v_n(\pt)$ obtained with the event plane method to the
      analogous measurements from ATLAS~\cite{Aad:2012bu} and
      CMS~\cite{Chatrchyan:2012ta} collaborations, as well as $v_2$
      measurements by STAR~\cite{Adams:2004bi}. Only statistical
      errors are shown.}
  \label{fig:vn_comparison}
  \end{center}
\end{figure}

Figure~\ref{fig:vn} shows unidentified charged particle $v_2$, $v_3$, and $v_4$ 
as a function of transverse momentum for different centrality classes. The 
difference between $v_2\{\rm EP\}$ and $v_2\{4\}$ for $\pt<7$~GeV/$c$ is 
predominantly due to flow fluctuations. The measured $v_2$ at $\pt>8$~GeV/$c$ 
is non-zero, positive and approximately constant, while its value increases 
from central to mid-peripheral collisions. The observed $v_2\{\rm EP\}$ at 
$\pt>10$~GeV/$c$ is fairly well described by extrapolation to the LHC 
energy~\cite{Horowitz:2011gd} of the WHDG model 
calculations~\cite{Wicks:2005gt} for $v_2$ of neutral pions including 
collisional and radiative energy loss of partons in a Bjorken-expanding 
medium~\cite{Horowitz:2011cv}. The coefficient $v_3$ exhibits a weak 
centrality dependence with a magnitude significantly smaller than that of 
$v_2$, except for the most central collisions. Unlike $v_3$, which originates 
entirely from fluctuations of the initial geometry of the system, $v_4$ has 
two contributions, which are probed by correlations with the $\Psi_{2}$ and 
$\Psi_{4}$ symmetry planes. The measured $v_{4/\Psi_{4}}\{\rm{EP}\}$ does not 
depend strongly on the collision centrality which points to a strong 
contribution from flow fluctuations. In contrast, $v_{4/\Psi_{2}}\{\rm{EP}\}$ 
shows a strong centrality dependence which is typical for correlations with 
respect to the true reaction plane. The difference between the two, 
indicative of flow fluctuations, persists at least up to $\pt= 8$~GeV/$c$.

Figure~\ref{fig:vn_comparison} compares our results obtained with the event 
plane method for 30-40\% centrality to the analogous measurements by 
ATLAS~\cite{Aad:2012bu} and CMS~\cite{Chatrchyan:2012ta} collaborations, and 
results obtained at RHIC by the STAR~\cite{Adams:2004bi} collaboration. An 
excellent agreement is observed between results from all three LHC 
experiments. $v_{2}(\pt)$ at top RHIC energy has a peak value about 10\% lower than at LHC although is very similar in shape.
 
\begin{figure}[t]   
  \begin{center}
  \includegraphics[keepaspectratio, width=0.7\columnwidth]{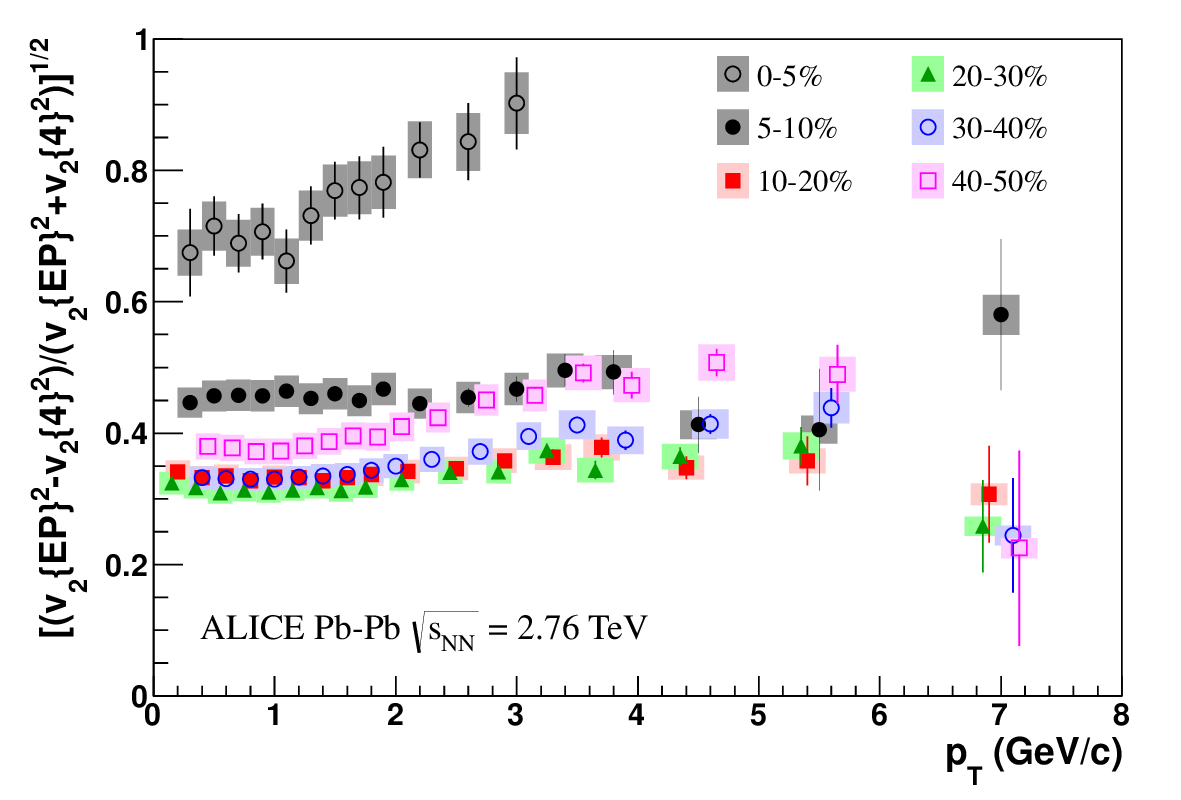}
  \caption{(color online) Relative event-by-event elliptic flow 
    fluctuations for unidentified charged particles versus transverse momentum 
    for different centrality classes. For clarity, the markers for 
    centrality classes $\ge 10\%$ are slightly shifted along the 
    horizontal axis. Error bars (shaded boxes) represent the 
    statistical (systematic) uncertainties.} 
  \label{fig:v2_ratio}
  \end{center}
\end{figure}

To investigate further the role of flow fluctuations at different 
transverse momenta we study the relative difference between 
$v_2\{\rm{EP}\}$ and $v_2\{4\}$, 
$[(v_2\{\mathrm{EP}\}^2-v_2\{4\}^2)/(v_2\{\mathrm{EP}\}^2+v_2\{4\}^2)]^{1/2}$, 
which for small non-flow is proportional to the relative flow 
fluctuations $\sigma_{v_2}/\langle v_{2} \rangle$~\cite{Voloshin:2008dg}. 
Figure~\ref{fig:v2_ratio} presents 
this quantity as a function of transverse momentum for various 
centrality classes. The relative flow fluctuations are minimal for mid-central 
collisions and become larger for peripheral and central collisions, similar to 
those observed at RHIC energies~\cite{Voloshin:2008dg}. It is remarkable that 
in the 5-30\% centrality range, relative flow 
fluctuations are within errors independent of momentum up to 
$\pt \sim 8$~GeV/$c$, far beyond the region where the flow magnitude is 
well described by hydrodynamic models ($\pt< 2-3$~GeV/$c$). This 
indicates a common origin for flow fluctuations, which are usually 
associated with fluctuations of the initial collision geometry, at 
least up to the regime where hard scattering and jet energy loss are 
expected to dominate. The ratio develops a momentum dependence, starting to 
increase at $\pt \sim 1.5$~GeV/$c$, for more peripheral collisions 
(30-50\%), and in most central collisions (0-5\%), where it is most 
pronounced. In both cases, the relative contribution of non-flow effects is 
expected to be the largest.

\begin{figure}[t]
  \begin{center}
  \includegraphics[keepaspectratio, width=0.7\columnwidth]{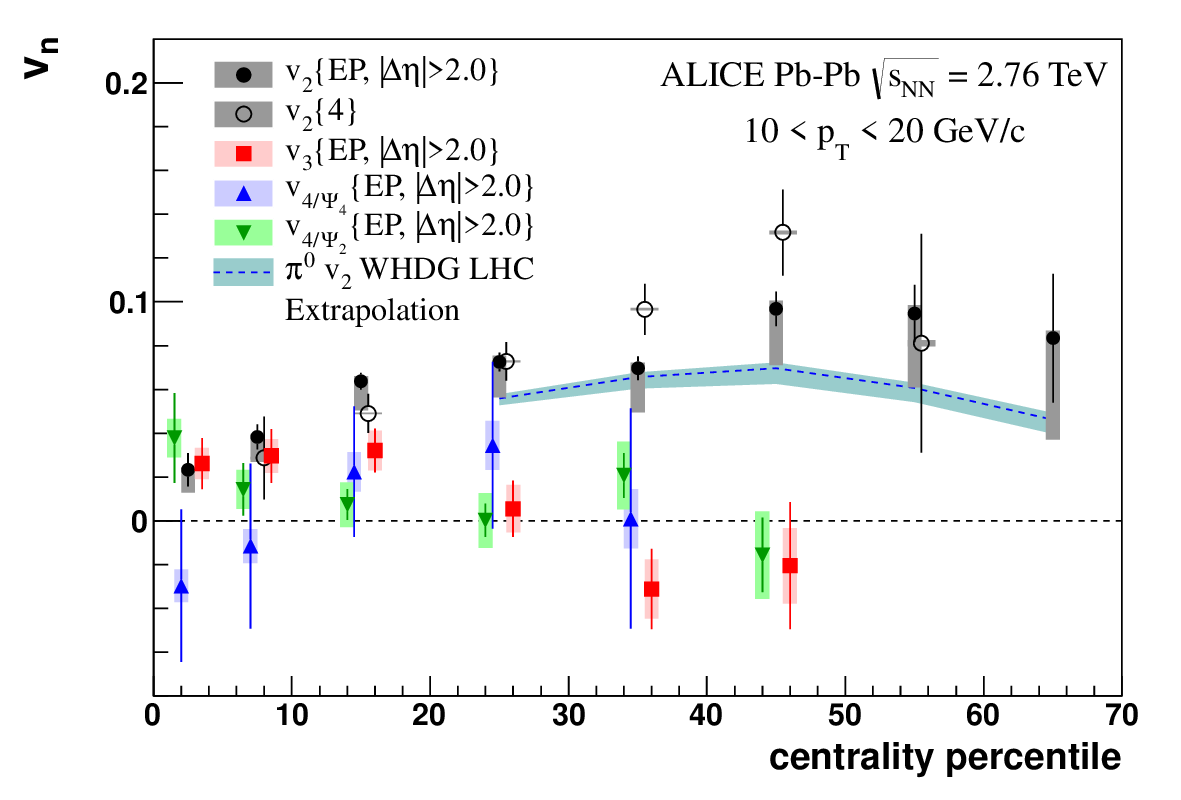}
  \caption{(color online) Unidentified charged particle $v_2$, $v_3$, and 
    $v_4$ integrated over the transverse momentum range $10<\pt<20$~GeV/$c$ 
    as a function of collision centrality, with the more central (peripheral) 
    collisions shown on the left-(right-)hand side, respectively. The dashed 
    line represents the WHDG model calculations for neutral 
    pions~\cite{Horowitz:2011cv} extrapolated to the LHC collision energy. 
    Error bars (shaded boxes) represent the statistical (systematic) 
    uncertainties.}
  \label{fig:vn_int_1020}
  \end{center}
\end{figure}

Figure~\ref{fig:vn_int_1020} shows unidentified charged particle
$v_2$, $v_3$, and $v_4$ averaged over $10<\pt<20$~GeV/$c$ as a
function of centrality. $v_2$ increases from central to peripheral
collisions. No significant difference between $v_2\{\rm EP\}$ 
and $v_2\{4\}$ results is observed, which might indicate that the fluctuations 
of the initial collision geometry become unimportant for $\pt > 10$ GeV/$c$. 
The centrality dependence of $v_3$ differs significantly from that of 
$v_2$. $v_4$ measured with 
respect to the second and fourth harmonic symmetry planes is consistent with 
zero within relatively large uncertainties. All these observations indicate 
that for $\pt>10$ GeV/$c$ the effect of fluctuations of the initial 
collision geometry might be very different compared to that at low and 
intermediate $\pt$.

\begin{figure}[t]
  \begin{center}
  \includegraphics[keepaspectratio, width=0.7\columnwidth]{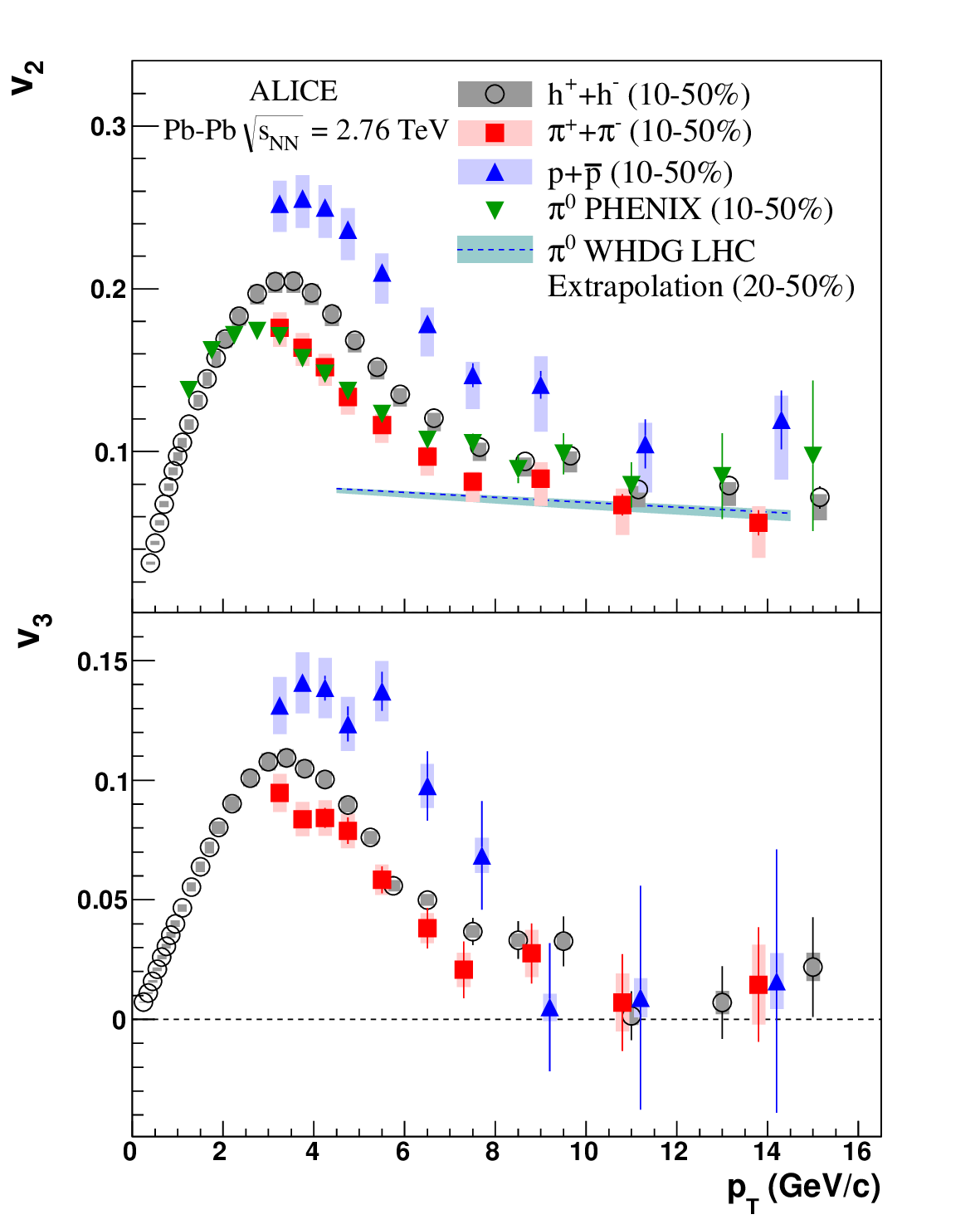}
  \caption{(color online) $v_2$ (top) and $v_3$ (bottom) of charged pion and 
    proton as a function of transverse momentum for 10-50\% centrality 
    class compared to unidentified charged particles results from the event 
    plane method. For clarity, the markers for $v_2$ and $v_3$ at 
    $\pt>8$~GeV/$c$ are 
    slightly shifted along the horizontal axis. PHENIX $\pi^0$ $v_2$ 
    measurements~\cite{Adare:2010sp} are also shown. The dashed line 
    represents the WHDG model calculations for neutral 
    pions~\cite{Horowitz:2011cv} extrapolated to the LHC collision energy 
    for the 20-50\% centrality range. Error bars (shaded boxes) represent the 
    statistical (systematic) uncertainties.}
  \label{fig:vn_pid}
  \end{center}
\end{figure}

Figure~\ref{fig:vn_pid} presents charged pion and proton $v_2$ and 
$v_3$ as a function of $\pt$ in the 10-50\% centrality range from the event 
plane method. The proton $v_2$ and $v_3$ are higher than that of pions 
out to $\pt=8$~GeV/$c$ where the uncertainties become large. This behavior 
is qualitatively consistent with a picture where particle production in this 
intermediate $\pt$ region includes interaction of jet fragments with bulk 
matter, e.g. as in model~\cite{Werner:2012v2}. The magnitude 
of the measured charged pion elliptic flow for 
$\pt>8$ GeV/$c$ is compatible with that for unidentified charged particles, 
and $\pi^0$ measured by PHENIX~\cite{Adare:2010sp} in Au-Au collisions 
at $\sqrt{s_{\rm{NN}}}=0.2$~TeV, and reproduced by the WHDG model 
calculations for $v_2$ of neutral pions~\cite{Horowitz:2011cv}.

In summary, we have presented elliptic, triangular, and quadrangular flow 
coefficients measured by the ALICE collaboration in Pb-Pb collisions at 
$\snn=2.76$~TeV over a broad range of transverse momentum.
For $\pt>8$~GeV/$c$, we find that the unidentified charged 
particle $v_2$ and $v_3$ are finite, positive and only weakly dependent on 
transverse momentum, while $v_4$ is consistent with zero within rather large 
statistical and systematic uncertainties. The observed 
difference in the centrality dependence of $v_{4/\Psi_{4}}$ and $v_{4/\Psi_{2}}$,
and the results on $v_2$ obtained with the event plane and four-particle 
cumulant methods indicate that the effect of flow fluctuations extends at 
least up to $\pt= 8$~GeV/$c$ and does not change significantly in magnitude. It 
shows that the effect of fluctuations of the initial collision geometry on 
particle production is similar at low and intermediate $\pt$ regions, which are 
considered to be dominated by hydrodynamical flow and quark coalescence, 
respectively. For $\pt>10$ GeV/$c$, where particle production is dominated by 
fragmentation of hard partons, the response to fluctuations of the initial 
collision geometry might be different, but more data is needed to study this 
regime in more detail. The pion $v_2$ at LHC energies is very close to that 
measured at RHIC out to $\pt=16$~GeV/$c$ and is reproduced by WHDG model 
calculations for $\pt >8$ GeV/$c$. The proton $v_2$ and $v_3$ are finite, 
positive, and have a larger magnitude than that of the pion for 
$\pt<8$~GeV/$c$, indicating that the particle type dependence, which is 
typical at low $\pt$, persists out to high transverse momenta.

\newenvironment{acknowledgement}{\relax}{\relax}
\begin{acknowledgement}
  \section*{Acknowledgements}
  The ALICE collaboration would like to thank all its engineers and technicians for their invaluable contributions to the construction of the experiment and the CERN accelerator teams for the outstanding performance of the LHC complex.
\\
The ALICE collaboration acknowledges the following funding agencies for their support in building and
running the ALICE detector:
 \\
Calouste Gulbenkian Foundation from Lisbon and Swiss Fonds Kidagan, Armenia;
 \\
Conselho Nacional de Desenvolvimento Cient\'{\i}fico e Tecnol\'{o}gico (CNPq), Financiadora de Estudos e Projetos (FINEP),
Funda\c{c}\~{a}o de Amparo \`{a} Pesquisa do Estado de S\~{a}o Paulo (FAPESP);
 \\
National Natural Science Foundation of China (NSFC), the Chinese Ministry of Education (CMOE)
and the Ministry of Science and Technology of China (MSTC);
 \\
Ministry of Education and Youth of the Czech Republic;
 \\
Danish Natural Science Research Council, the Carlsberg Foundation and the Danish National Research Foundation;
 \\
The European Research Council under the European Community's Seventh Framework Programme;
 \\
Helsinki Institute of Physics and the Academy of Finland;
 \\
French CNRS-IN2P3, the `Region Pays de Loire', `Region Alsace', `Region Auvergne' and CEA, France;
 \\
German BMBF and the Helmholtz Association;
\\
General Secretariat for Research and Technology, Ministry of
Development, Greece;
\\
Hungarian OTKA and National Office for Research and Technology (NKTH);
 \\
Department of Atomic Energy and Department of Science and Technology of the Government of India;
 \\
Istituto Nazionale di Fisica Nucleare (INFN) of Italy;
 \\
MEXT Grant-in-Aid for Specially Promoted Research, Ja\-pan;
 \\
Joint Institute for Nuclear Research, Dubna;
 \\
National Research Foundation of Korea (NRF);
 \\
CONACYT, DGAPA, M\'{e}xico, ALFA-EC and the HELEN Program (High-Energy physics Latin-American--European Network);
 \\
Stichting voor Fundamenteel Onderzoek der Materie (FOM) and the Nederlandse Organisatie voor Wetenschappelijk Onderzoek (NWO), Netherlands;
 \\
Research Council of Norway (NFR);
 \\
Polish Ministry of Science and Higher Education;
 \\
National Authority for Scientific Research - NASR (Autoritatea Na\c{t}ional\u{a} pentru Cercetare \c{S}tiin\c{t}ific\u{a} - ANCS);
 \\
Federal Agency of Science of the Ministry of Education and Science of Russian Federation, International Science and
Technology Center, Russian Academy of Sciences, Russian Federal Agency of Atomic Energy, Russian Federal Agency for Science and Innovations and CERN-INTAS;
 \\
Ministry of Education of Slovakia;
 \\
Department of Science and Technology, South Africa;
 \\
CIEMAT, EELA, Ministerio de Educaci\'{o}n y Ciencia of Spain, Xunta de Galicia (Conseller\'{\i}a de Educaci\'{o}n),
CEA\-DEN, Cubaenerg\'{\i}a, Cuba, and IAEA (International Atomic Energy Agency);
 \\
Swedish Research Council (VR) and Knut $\&$ Alice Wallenberg
Foundation (KAW);
 \\
Ukraine Ministry of Education and Science;
 \\
United Kingdom Science and Technology Facilities Council (STFC);
 \\
The United States Department of Energy, the United States National
Science Foundation, the State of Texas, and the State of Ohio.
\end{acknowledgement}
%


\newpage

\appendix
\section{The ALICE Collaboration}
\label{app:collab}

\begingroup
\small
\begin{flushleft}
B. Abelev\Irefn{a1234}\And
J. Adam\Irefn{a1274}\And
D. Adamov\'{a}\Irefn{a1283}\And
A.M. Adare\Irefn{a1260}\And
M.M. Aggarwal\Irefn{a1157}\And
G. Aglieri Rinella\Irefn{a1192}\And
A.G. Agocs\Irefn{a1143}\And
A. Agostinelli\Irefn{a1132}\And
S. Aguilar Salazar\Irefn{a1247}\And
Z. Ahammed\Irefn{a1225}
% ...
\renewcommand\labelenumi{\textsuperscript{\theenumi}~}
\section*{Affiliation notes}
\renewcommand\theenumi{\roman{enumi}}
\begin{Authlist}
\item \Adef{a1133}Also at: Sezione INFN, Bologna, Italy
\item \Adef{a1215}Also at: Gangneung-Wonju National University, Gangneung, South Korea
\item \Adef{a1290}Also at: Dipartimento di Fisica `E.R.~Caianiello' dell'Universit\`{a} and Gruppo Collegato INFN, Salerno, Italy
\item \Adef{a1139}Also at: Institute of Space Sciences (ISS), Bucharest, Romania
\item \Adef{a1192}Also at: European Organization for Nuclear Research (CERN), Geneva, Switzerland
\item \Adef{a1227}Now at: Fachhochschule K\"{o}ln, K\"{o}ln, Germany
\item \Adef{a1230}Also at: Institute of Experimental Physics, Slovak Academy of Sciences, Ko\v{s}ice, Slovakia
\item \Adef{a1160}Also at: Laboratoire de Physique Corpusculaire (LPC), Clermont Universit\'{e}, Universit\'{e} Blaise Pascal, CNRS--IN2P3, Clermont-Ferrand, France
\item \Adef{a1168}Now at: The Henryk Niewodniczanski Institute of Nuclear Physics, Polish Academy of Sciences, Cracow, Poland
\item \Adef{a1176}Now at: Research Division and ExtreMe Matter Institute EMMI, GSI Helmholtzzentrum f\"ur Schwerionenforschung, Darmstadt, Germany
\item \Adef{a1335}Also at: Centro Fermi -- Centro Studi e Ricerche e Museo Storico della Fisica ``Enrico Fermi'', Rome, Italy
\item \Adef{a1242}Also at: Centro de Investigaciones Energ\'{e}ticas Medioambientales y Tecnol\'{o}gicas (CIEMAT), Madrid, Spain
\item \Adef{a1256}Now at: Institut f\"{u}r Kernphysik, Westf\"{a}lische Wilhelms-Universit\"{a}t M\"{u}nster, M\"{u}nster, Germany
\end{Authlist}
\section*{Collaboration Institutes}
\renewcommand\theenumi{\arabic{enumi}~}
\begin{Authlist}
\item \Idef{a21360}Eberhard Karls Universit\"{a}t T\"{u}bingen, T\"{u}bingen, Germany
\item \Idef{a1320}Nikhef, National Institute for Subatomic Physics and Institute for Subatomic Physics of Utrecht University, Utrecht, Netherlands
\item \Idef{a1270}Dipartimento di Fisica dell'Universit\`{a} and Sezione INFN, Padova, Italy
\item \Idef{a15782}COMSATS Institute of Information Technology (CIIT), Islamabad, Pakistan
\item \Idef{a1160}Laboratoire de Physique Corpusculaire (LPC), Clermont Universit\'{e}, Universit\'{e} Blaise Pascal, CNRS--IN2P3, Clermont-Ferrand, France
\item \Idef{a1301}Yonsei University, Seoul, South Korea
\item \Idef{a1332}Yerevan Physics Institute, Yerevan, Armenia
\item \Idef{a1218}Scientific Research Technological Institute of Instrument Engineering, Kharkov, Ukraine
\item \Idef{a1187}Laboratori Nazionali di Frascati, INFN, Frascati, Italy
\item \Idef{a1125}Lawrence Berkeley National Laboratory, Berkeley, California, United States
\item \Idef{a1200}Physikalisches Institut, Ruprecht-Karls-Universit\"{a}t Heidelberg, Heidelberg, Germany
\item \Idef{a1224}Saha Institute of Nuclear Physics, Kolkata, India
\item \Idef{a1229}Faculty of Science, P.J.~\v{S}af\'{a}rik University, Ko\v{s}ice, Slovakia
\item \Idef{a1168}The Henryk Niewodniczanski Institute of Nuclear Physics, Polish Academy of Sciences, Cracow, Poland
\item \Idef{a1334}Rudjer Bo\v{s}kovi\'{c} Institute, Zagreb, Croatia
\item \Idef{a1251}Moscow Engineering Physics Institute, Moscow, Russia
\item \Idef{a1232}Laboratori Nazionali di Legnaro, INFN, Legnaro, Italy
\item \Idef{a1122}Faculty of Engineering, Bergen University College, Bergen, Norway
\item \Idef{a36377}Nuclear Physics Group, STFC Daresbury Laboratory, Daresbury, United Kingdom
\item \Idef{a1185}Institut f\"{u}r Kernphysik, Johann Wolfgang Goethe-Universit\"{a}t Frankfurt, Frankfurt, Germany
\item \Idef{a1268}Department of Physics, University of Oslo, Oslo, Norway
\item \Idef{a1310}University of Tokyo, Tokyo, Japan
\item \Idef{a1275}Institute of Physics, Academy of Sciences of the Czech Republic, Prague, Czech Republic
\item \Idef{a1312}Dipartimento di Fisica Sperimentale dell'Universit\`{a} and Sezione INFN, Turin, Italy
\item \Idef{a1192}European Organization for Nuclear Research (CERN), Geneva, Switzerland
\item \Idef{a1106}Department of Physics Aligarh Muslim University, Aligarh, India
\item \Idef{a1146}Sezione INFN, Cagliari, Italy
\item \Idef{a1250}Institute for Theoretical and Experimental Physics, Moscow, Russia
\item \Idef{a1212}Helsinki Institute of Physics (HIP) and University of Jyv\"{a}skyl\"{a}, Jyv\"{a}skyl\"{a}, Finland
\item \Idef{a1315}Dipartimento di Fisica dell'Universit\`{a} and Sezione INFN, Trieste, Italy
\item \Idef{a1130}School of Physics and Astronomy, University of Birmingham, Birmingham, United Kingdom
\item \Idef{a1109}Nikhef, National Institute for Subatomic Physics, Amsterdam, Netherlands
\item \Idef{a1338}Secci\'{o}n F\'{\i}sica, Departamento de Ciencias, Pontificia Universidad Cat\'{o}lica del Per\'{u}, Lima, Peru
\item \Idef{a20958}Gauhati University, Department of Physics, Guwahati, India
\item \Idef{a1266}Institut de Physique Nucl\'{e}aire d'Orsay (IPNO), Universit\'{e} Paris-Sud, CNRS-IN2P3, Orsay, France
\item \Idef{a1139}Institute of Space Sciences (ISS), Bucharest, Romania
\item \Idef{a1209}Physics Department, University of Jammu, Jammu, India
\item \Idef{a1182}Joint Institute for Nuclear Research (JINR), Dubna, Russia
\item \Idef{a1239}Universit\'{e} de Lyon, Universit\'{e} Lyon 1, CNRS/IN2P3, IPN-Lyon, Villeurbanne, France
\item \Idef{a1292}California Polytechnic State University, San Luis Obispo, California, United States
\item \Idef{a1262}Budker Institute for Nuclear Physics, Novosibirsk, Russia
\item \Idef{a1290}Dipartimento di Fisica `E.R.~Caianiello' dell'Universit\`{a} and Gruppo Collegato INFN, Salerno, Italy
\item \Idef{a1285}Dipartimento di Fisica dell'Universit\`{a} `La Sapienza' and Sezione INFN, Rome, Italy
\item \Idef{a1222}University of Tennessee, Knoxville, Tennessee, United States
\item \Idef{a1277}Institute for High Energy Physics, Protvino, Russia
\item \Idef{a1296}Universidade de S\~{a}o Paulo (USP), S\~{a}o Paulo, Brazil
\item \Idef{a1132}Dipartimento di Fisica dell'Universit\`{a} and Sezione INFN, Bologna, Italy
\item \Idef{a1244}Centro de Investigaci\'{o}n y de Estudios Avanzados (CINVESTAV), Mexico City and M\'{e}rida, Mexico
\item \Idef{a1140}National Institute for Physics and Nuclear Engineering, Bucharest, Romania
\item \Idef{a14939}Centre de Calcul de l'IN2P3, Villeurbanne, France
\item \Idef{a1184}Frankfurt Institute for Advanced Studies, Johann Wolfgang Goethe-Universit\"{a}t Frankfurt, Frankfurt, Germany
\item \Idef{a1225}Variable Energy Cyclotron Centre, Kolkata, India
\item \Idef{a1207}Physics Department, University of Rajasthan, Jaipur, India
\item \Idef{a1215}Gangneung-Wonju National University, Gangneung, South Korea
\item \Idef{a1242}Centro de Investigaciones Energ\'{e}ticas Medioambientales y Tecnol\'{o}gicas (CIEMAT), Madrid, Spain
\item \Idef{a1194}Laboratoire de Physique Subatomique et de Cosmologie (LPSC), Universit\'{e} Joseph Fourier, CNRS-IN2P3, Institut Polytechnique de Grenoble, Grenoble, France
\item \Idef{a27399}Institut f\"{u}r Informatik, Johann Wolfgang Goethe-Universit\"{a}t Frankfurt, Frankfurt, Germany
\item \Idef{a1306}V.~Fock Institute for Physics, St. Petersburg State University, St. Petersburg, Russia
\item \Idef{a1162}Department of Physics, Ohio State University, Columbus, Ohio, United States
\item \Idef{a1335}Centro Fermi -- Centro Studi e Ricerche e Museo Storico della Fisica ``Enrico Fermi'', Rome, Italy
\item \Idef{a1145}Dipartimento di Fisica dell'Universit\`{a} and Sezione INFN, Cagliari, Italy
\item \Idef{a1318}University of Tsukuba, Tsukuba, Japan
\item \Idef{a1136}Faculty of Mathematics, Physics and Informatics, Comenius University, Bratislava, Slovakia
\item \Idef{a1199}Kirchhoff-Institut f\"{u}r Physik, Ruprecht-Karls-Universit\"{a}t Heidelberg, Heidelberg, Germany
\item \Idef{a1300}Department of Physics, Sejong University, Seoul, South Korea
\item \Idef{a1329}Hua-Zhong Normal University, Wuhan, China
\item \Idef{a1173}Universidad Aut\'{o}noma de Sinaloa, Culiac\'{a}n, Mexico
\item \Idef{a1327}Zentrum f\"{u}r Technologietransfer und Telekommunikation (ZTT), Fachhochschule Worms, Worms, Germany
\item \Idef{a1205}University of Houston, Houston, Texas, United States
\item \Idef{a36378}Indian Institute of Technology Indore (IIT), Indore, India
\item \Idef{a1197}Centro de Aplicaciones Tecnol\'{o}gicas y Desarrollo Nuclear (CEADEN), Havana, Cuba
\item \Idef{a1246}Instituto de Ciencias Nucleares, Universidad Nacional Aut\'{o}noma de M\'{e}xico, Mexico City, Mexico
\item \Idef{a1176}Research Division and ExtreMe Matter Institute EMMI, GSI Helmholtzzentrum f\"ur Schwerionenforschung, Darmstadt, Germany
\item \Idef{a1155}Sezione INFN, Catania, Italy
\item \Idef{a1274}Faculty of Nuclear Sciences and Physical Engineering, Czech Technical University in Prague, Prague, Czech Republic
\item \Idef{a1281}Pusan National University, Pusan, South Korea
\item \Idef{a1279}Benem\'{e}rita Universidad Aut\'{o}noma de Puebla, Puebla, Mexico
\item \Idef{a1258}SUBATECH, Ecole des Mines de Nantes, Universit\'{e} de Nantes, CNRS-IN2P3, Nantes, France
\item \Idef{a17361}The University of Texas at Austin, Physics Department, Austin, TX, United States
\item \Idef{a1249}Institute for Nuclear Research, Academy of Sciences, Moscow, Russia
\item \Idef{a1154}Dipartimento di Fisica e Astronomia dell'Universit\`{a} and Sezione INFN, Catania, Italy
\item \Idef{a1322}Soltan Institute for Nuclear Studies, Warsaw, Poland
\item \Idef{a1121}Department of Physics and Technology, University of Bergen, Bergen, Norway
\item \Idef{a1230}Institute of Experimental Physics, Slovak Academy of Sciences, Ko\v{s}ice, Slovakia
\item \Idef{a20954}Korea Institute of Science and Technology Information, Daejeon, South Korea
 \item \Idef{a1177}Institut f\"{u}r Kernphysik, Technische Universit\"{a}t Darmstadt, Darmstadt, Germany
\item \Idef{a1157}Physics Department, Panjab University, Chandigarh, India
\item \Idef{a1189}Petersburg Nuclear Physics Institute, Gatchina, Russia
\item \Idef{a1152}Physics Department, University of Cape Town, iThemba LABS, Cape Town, South Africa
\item \Idef{a1115}Sezione INFN, Bari, Italy
\item \Idef{a1127}Institute of Physics, Bhubaneswar, India
\item \Idef{a1313}Sezione INFN, Turin, Italy
\item \Idef{a1298}Russian Federal Nuclear Center (VNIIEF), Sarov, Russia
\item \Idef{a1179}Wayne State University, Detroit, Michigan, United States
\item \Idef{a1323}Warsaw University of Technology, Warsaw, Poland
\item \Idef{a1325}Purdue University, West Lafayette, Indiana, United States
\item \Idef{a1114}Dipartimento Interateneo di Fisica `M.~Merlin' and Sezione INFN, Bari, Italy
\item \Idef{a1308}Institut Pluridisciplinaire Hubert Curien (IPHC), Universit\'{e} de Strasbourg, CNRS-IN2P3, Strasbourg, France
\item \Idef{a1256}Institut f\"{u}r Kernphysik, Westf\"{a}lische Wilhelms-Universit\"{a}t M\"{u}nster, M\"{u}nster, Germany
\item \Idef{a1286}Sezione INFN, Rome, Italy
\item \Idef{a1112}Physics Department, University of Athens, Athens, Greece
\item \Idef{a1283}Nuclear Physics Institute, Academy of Sciences of the Czech Republic, \v{R}e\v{z} u Prahy, Czech Republic
\item \Idef{a1247}Instituto de F\'{\i}sica, Universidad Nacional Aut\'{o}noma de M\'{e}xico, Mexico City, Mexico
\item \Idef{a1271}Sezione INFN, Padova, Italy
\item \Idef{a1220}Bogolyubov Institute for Theoretical Physics, Kiev, Ukraine
\item \Idef{a1294}Departamento de F\'{\i}sica de Part\'{\i}culas and IGFAE, Universidad de Santiago de Compostela, Santiago de Compostela, Spain
\item \Idef{a1237}Division of Experimental High Energy Physics, University of Lund, Lund, Sweden
\item \Idef{a1316}Sezione INFN, Trieste, Italy
\item \Idef{a1203}Hiroshima University, Hiroshima, Japan
\item \Idef{a20959}Bose Institute, Department of Physics and Centre for Astroparticle Physics and Space Science (CAPSS), Kolkata, India
\item \Idef{a1264}Oak Ridge National Laboratory, Oak Ridge, Tennessee, United States
\item \Idef{a1170}Physics Department, Creighton University, Omaha, Nebraska, United States
\item \Idef{a1149}Universidade Estadual de Campinas (UNICAMP), Campinas, Brazil
\item \Idef{a1304}Technical University of Split FESB, Split, Croatia
\item \Idef{a1252}Russian Research Centre Kurchatov Institute, Moscow, Russia
\item \Idef{a1165}Niels Bohr Institute, University of Copenhagen, Copenhagen, Denmark
\item \Idef{a1133}Sezione INFN, Bologna, Italy
\item \Idef{a1103}Dipartimento di Scienze e Tecnologie Avanzate dell'Universit\`{a} del Piemonte Orientale and Gruppo Collegato INFN, Alessandria, Italy
\item \Idef{a1260}Yale University, New Haven, Connecticut, United States
\item \Idef{a17347}Chicago State University, Chicago, United States
\item \Idef{a1227}Fachhochschule K\"{o}ln, K\"{o}ln, Germany
\item \Idef{a1118}China Institute of Atomic Energy, Beijing, China
\item \Idef{a1288}Commissariat \`{a} l'Energie Atomique, IRFU, Saclay, France
\item \Idef{a1254}Indian Institute of Technology, Mumbai, India
\item \Idef{a1234}Lawrence Livermore National Laboratory, Livermore, California, United States
\item \Idef{a1143}KFKI Research Institute for Particle and Nuclear Physics, Hungarian Academy of Sciences, Budapest, Hungary
\end{Authlist}
\endgroup


\end{document}